\documentstyle[preprint,pre,aps]{revtex}
\begin{document}
\draft
\title{Period Doublings
       in Coupled Parametrically Forced Damped Pendulums
}
\author{Sang-Yoon Kim
\footnote{Permanent address: Department of Physics, Kangwon National
          University, Chunchon, Kangwon-Do 200-701, Korea.
          Electronic address: sykim@cc.kangwon.ac.kr}
       }
\address{
School of Physics, Georgia Institute of Technology,
Atlanta, Georgia 30332-0430
}
\author{Kijin Lee}
\address{Department of Physics, Kangwon National
          University, Chunchon, Kangwon-Do 200-701, Korea}
\maketitle

\begin{abstract}
We study period doublings in $N$ $(N=2,3,4, \dots)$ coupled
parametrically forced damped pendulums by varying $A$ (the
amplitude of the external
driving force) and $c$ (the strength of coupling). With increasing
$A$, the stationary point undergoes multiple period-doubling
transitions to chaos. We
first investigate the two-coupled case with $N=2$. For each
period-doubling transition to chaos, the critical set consists of
an infinity of critical line segments and the zero-coupling critical
point lying on the line
$A=A^*_i$ in the $A-c$ plane, where $A^*_i$ is the $i$th transition
point for the uncoupled case. We find three kinds of critical
behaviors, depending on the position on the critical set. They are
the same as those for the coupled one-dimensional maps. Finally, the
results of the $N=2$ case are extended to many-coupled cases with
$N \geq 3$, in which the critical behaviors
depend on the range of coupling.

\end{abstract}


%
%

\narrowtext

\section{Introduction}
\label{sec:Int}

In recent years, much attention has been paid to coupled nonlinear
oscillators. Such coupled oscillators are used to model many
physical, chemical, and biological systems such as coupled p-n
junctions \cite{Jeffries},
Josephson-junction arrays \cite{Hadley}, the charge-density waves
\cite{Strogatz}, chemical-reaction systems\cite{Kuramoto}, and
biological-oscillation systems \cite{Winfree}. They are known to
exhibit period-doubling bifurcations (PDB's), saddle-node
bifurcations, Hopf bifurcations, chaos as well as pattern formation.

The coupled oscillators investigated in this paper are coupled
parametrically forced damped pendulums (PFDP's). For a single damped
pendulum, vertical oscillation of its support leads to a
time-periodic variation of
its natural frequency, and hence it is called a
PFDP \cite{Landau,Arnold}. This simple PFDP shows richness in its
dynamical behavior \cite{McLaughlin,Leven,Arneodo}.
One of the interesting behaviors is that with
increasing the amplitude $A$ of the vertical oscillation, the
stationary point undergoes multiple period-doubling transitions to
chaos, which have been found in our recent work \cite{Kim1}.

Here we study the critical behaviors of PDB's in $N$ $(N=2,3,4,\dots)$
coupled PFDP's. The ``coupling effect'' of the nature, strength, and
range of coupling on the critical behaviors are particularly
investigated. The coupled PFDP's exhibit multiple period-doubling
transitions to chaos. This is in contrast
to the case of the coupled one-dimensional (1D) maps, in which only
single period-doubling transition to chaos occurs \cite{KK,Kim2}.
We first consider the simplest coupled
case with $N=2$. For each period-doubling transition to chaos, the
critical set (set of the critical points) is composed of the
zero-coupling critical
point and an infinity of critical line segments lying on the line
$A=A^*_i$ in the $A-c$ plane, where $A$ is the amplitude of the
external driving force, $c$ a coupling parameter, and $A^*_i$ the
$i$th period-doubling transition point for the uncoupled case.
It is found that there exist three kinds of critical behaviors,
depending on the position on the critical set. These critical
behaviors are the same as those of the coupled one-dimensional (1D)
maps found by one of us and Kook \cite{Kim2}. The results
for the two-coupled case are also extended to many-coupled cases with
$N \geq 3$. It is found that the critical behaviors for the
many-coupled cases vary depending
on whether or not the coupling is global. In the extreme long-range
case of global coupling, in which each PFDP is coupled to all the
other ones with equal coupling strength, the critical behaviors are
the same as those for the two-coupled case,
irrespectively of $N$. However, for any other nonglobal-coupling cases
of shorter-range couplings, an important change occurs in the
stability diagram of $2^n$-periodic $(n=0,1,2,\dots)$ orbits in the
$A-c$ plane, and consequently the structure of the critical set
becomes different from that for the
global-coupling case. To the best of our knowledge, so far only the
authors of the paper \cite{Kook1} attempted to study the critical
behaviors of PDB's in coupled
oscillators. However, only the critical behaviors near the
zero-coupling critical point were
considered, because the existence of an infinity of additional
critical line segments in the coupled 1D maps was not known at that
time. Moreover, any explicit numerical values of the scaling factors
for the zero-coupling case were not obtained \cite{Kook2}.

This paper is organized as follows. We first introduce two coupled
PFDP's and discuss their general properties in Sec.~\ref{sec:TC}.
Stability of periodic orbits, bifurcations, and Lyapunov exponents are
also discussed there. In Sec.~\ref{sec:CBTC}, the critical behaviors
of PDB's for the
two-coupled case are studied by varying two parameters $A$ and $c$.
We also extend the results of the two-coupled case
to many coupled PFDP's in Sec.~\ref{sec:CBMC}. Finally, a summary is
given in Sec.~\ref{sec:Sum}.

\section{Stability of Periodic Orbits, Bifurcations, and Lyapunov
         exponents in two coupled PFDP's}
\label{sec:TC}

In this section, we first discuss stability of period orbits in the
Poincar{\'e} map of the two coupled PFDP's, using the Floquet theory.
Bifurcations associated with the stability and Lyapunov exponents are
then discussed.

Consider a system consisting of two identical PFDP's
coupled symmetrically:
\begin{mathletters}
\begin{eqnarray}
{\ddot x_1} &=& f(x_1,{\dot x}_1,t) + g(x_1,x_2),  \\
{\ddot x_2} &=& f(x_2,{\dot x}_2,t) + g(x_2,x_1).
\end{eqnarray}
\label{eq:TPFDP1}
\end{mathletters}
where
\begin{equation}
f(x,{\dot x},t)=-2 \pi \gamma {\dot x} - 2 \pi
(\omega_0^2-A \cos 2 \pi t) \sin 2 \pi x
\label{eq:fftn}
\end{equation}
and $g(x_1,x_2)$ is a coupling function, obeying the condition
$g(x,x)=0$ for all $x$. Here $x$ is the angular position, $\gamma$
the damping coefficient, $\omega_0$ the natural frequency of the
pendulum, $A$ the amplitude of the external driving force of period
one, and we consider the coupling function $g(x_1,x_2)$ of the form,
\begin{equation}
g(x_1,x_2)= {c \over 2} [u(x_2) - u(x_1)]
\label{eq:CFTN}
\end{equation}
where $u(x)$ is a function of one variable, and $c$ a coupling
parameter.

The two second-order ordinary differential equations (\ref{eq:TPFDP1})
are reduced to four first-order ordinary differential equations:
\begin{mathletters}
\begin{eqnarray}
{\dot x_1} &=& y_1, \\
{\dot y_1} &=& f(x_1,y_1,t) + g(x_1,x_2), \\
{\dot x_2} &=& y_2, \\
{\dot y_2} &=& f(x_2,y_2,t) + g(x_2,x_1).
\end{eqnarray}
\label{eq:TPFDP2}
\end{mathletters}
Consider an initial point ${\bf z}(0) [\equiv (z_1(0),z_2(0))]$, where
$z_i = (x_i,y_i)$ $(i=1,2)$. Then, its Poincar{\'e} maps can be
computed by sampling the points ${\bf z}(m)$ at the discrete time $m$,
where $m=1,2,3, \dots$~. We call
the transformation ${\bf z}(m) \rightarrow {\bf z}(m+1)$ the
Poincar{\'e} (time-$1$) map, and write ${\bf z}(m+1) = P({\bf z}(m))$.

The four-dimensional (4D) Poincar{\'e} map $P$ has an exchange
symmetry such that
\begin{equation}
S_1 P S_1({\bf z}) = P({\bf z})~{\rm for ~all~}{\bf z},
\end{equation}
where $S_1(z_1,z_2) = (z_2,z_1)$. The set of all points, which are
invariant under the exchange of coordinates $S_1$, forms a synchronous
plane on which
$x_1=x_2$ and $y_1=y_2$.
An orbit is called a(n) (in-phase) synchronous orbit if it lies on the
synchronous plane, i.e., it satisfies
\begin{equation}
x_1(m)=x_2(m) \equiv x^*(m),~~y_1(m)=y_2(m) \equiv y^*(m)
~{\rm for~all~}m.
\end{equation}
Otherwise, it is called an (out-of-phase) asynchronous orbit.
Here we study only the synchronous orbits.
They can be easily found from the uncoupled PFDP, because the coupling
function satisfies $g(x^*,x^*)=0$. Note also that for the cases of
these synchronous orbits, the 4D Poincar{\'e} map $P$ also has the
inversion symmetry such that
\begin{equation}
S_2 P S_2 ({\bf z}) = P({\bf z})~{\rm for ~all~}{\bf z},
\end{equation}
where $S_2({\bf z})= - {\bf z}$.
If a synchronous orbit $\{ {\bf z}(m) \}$ of $P$ is invariant under
$S_2$, it is called a symmetric orbit. Otherwise, it is called an
asymmetric orbit and has its ``conjugate'' orbits
$S_2 \{ {\bf z}(m) \}$.

We now study the stability of a synchronous periodic orbit with
period $q$ such that $P^q ({\bf z}(0)) ={\bf z}(0)$ but
$P^j ({\bf z}(0)) \neq {\bf z}(0)$
for $1 \leq j \leq q-1$. Here $P^k$ means the $k$-times iterated map.
The linear stability of the $q$-periodic orbit is determined from the
linearized-map matrix $DP^q({\bf z}(0))$ of $P^q$ at an orbit point
${\bf z}(0)$. Using the Floquet theory \cite{Lefschetz1}, the matrix
$DP^q$ can be obtained by integrating the linearized differential
equations for small perturbations as follows.

Stability analysis of an orbit can be conveniently carried out
in a set of new coordinates $(X_1,Y_1,X_2,Y_2)$ defined by
\begin{mathletters}
\begin{eqnarray}
X_1&=&{{(x_1+x_2)} \over 2},~ Y_1={{(y_1+y_2)} \over 2}, \\
X_2&=&{{(x_1-x_2)} \over 2},~ Y_2={{(y_1-y_2)} \over 2}.
\end{eqnarray}
\label{eq:NC}
\end{mathletters}
Here the first and second pairs of coordinates $Z_1$ and $Z_2$,
defined by $Z_i \equiv (X_i,Y_i)$ $(i=1,2)$, correspond to the
synchronous and asynchronous modes of the orbit, respectively. For
example, for a synchronous orbit $Z_1=(x^*,y^*)$ and $Z_2=(0,0)$,
while for an asynchronous orbit, $Z_2 \neq (0,0)$. Hereafter, we
will call $Z_1$ and $Z_2$ the synchronous and asynchronous modes of
the orbit, respectively.

Let ${\bf Z} (t)$ $[\equiv (Z_1,Z_2)]$ be a solution lying on
the closed orbit corresponding to a synchronous $q-$periodic orbit
with ${\bf Z}(t)={\bf Z}(t+q)$. In order to study the stability of
the synchronous closed orbit, we consider an infinitesimal
perturbation $\delta {\bf Z}$ $[\equiv (\delta Z_1, \delta Z_2)]$ to
the orbit. Note that $\delta Z_1$ and $\delta Z_2$ are the synchronous
and asynchronous modes of the perturbation to the synchronous orbit,
respectively. Linearizing the ordinary differential equations
(\ref{eq:TPFDP2}) (expressed in terms of the new
coordinates) about the orbit, we obtain
\begin{equation}
 \left( \begin{array}{c}
         \delta {\dot Z_1}  \\
         \delta {\dot Z_2}
      \end{array}
      \right)
      = J(t)
      \left( \begin{array}{c}
         \delta Z_1  \\
         \delta Z_2
      \end{array}
      \right),
\label{eq:LEQ}
\end{equation}
where
\begin{equation}
J = \left(
   \begin{array} {cc}
   J_1 & {\bf 0} \\
   {\bf 0} & J_2
   \end{array}
   \right).
\label{eq:J}
\end{equation}
Here ${\bf 0}$ is the $2 \times 2$ null matrix, and
\begin{eqnarray}
 J_1(t) &=&
 \left( \begin{array} {cc}
        0 & 1 \\
        f_1(x^*,{\dot x}^*,t) & f_2(x^*,{\dot x}^*,t)
        \end{array}
 \right),      \\
 J_2(t) &=&
 \left( \begin{array} {cc}
        0 & 1 \\
        f_1(x^*,{\dot x}^*,t)- 2 G(x^*) & f_2(x^*,{\dot x}^*,t)
        \end{array}
 \right),
\end{eqnarray}
where
\begin{eqnarray}
f_1(x^*,{\dot x}^*,t) &\equiv&
      { {\partial f(x^*,{\dot x}^*,t)} \over {\partial x^*} }
      =-4 \pi^2 (\omega_0^2 - A \cos 2 \pi t) \cos 2 \pi x^*(t), \\
f_2(x^*,{\dot x}^*,t) &\equiv&
      { {\partial f(x^*,{\dot x}^*,t)} \over {\partial {\dot x}^*} }
      =-2 \pi \gamma,
\end{eqnarray}
and
\begin{equation}
 G(x^*) \equiv \displaystyle{ \left.
      { {\partial g(x_1,x_2)} \over {\partial x_2} }
      \right|_{x_1=x_2=x^*} } = { c \over 2} u'(x^*).
\end{equation}
Here the prime denotes the differentiation.

Since the $4 \times 4$ matrix $J$ of Eq.~(\ref{eq:J}) is decomposed
into two $2 \times 2$ submatrices $J_1$ and $J_2$, Eq.~(\ref{eq:LEQ})
is reduced to two independent equations,
\begin{equation}
\delta {\dot Z}_i = J_i(t) \delta Z_i~{\rm for~} i=1, 2.
\label{eq:TLEQ}
\end{equation}
That is, $\delta Z_1$ (synchronous-mode perturbation) and $\delta Z_2$
(asynchronous-mode perturbation) become decoupled for the case of a
synchronous orbit. Note also that each $J_i$ $(i=1,2)$ is a
$q$-periodic matrix. Let $W_i(t)=(w^{(1)}_i(t),w^{(2)}_i(t))$ be a
fundamental solution
matrix with $W_i(0) = I$. Here $w^{(1)}_i(t)$ and $w^{(2)}_i(t)$ are
two independent solutions expressed in column vector forms, and $I$
is the $2 \times 2$ unit matrix. Then a general solution of the
$q$-periodic system has the following form
\begin{equation}
 \delta Z_i(t) = W_i (t) \delta Z_i(0), ~W_i(0)=I.
\label{eq:FSM}
\end{equation}
Substitution of Eq.~(\ref{eq:FSM}) into Eq.~(\ref{eq:TLEQ}) leads to
an initial-value problem to determine $W_i(t)$,
\begin{equation}
{\dot W}_i(t) = J_i(t) W_i(t),~W_i(0)=I.
\label{eq:WEQ}
\end{equation}
In this system of new coordinates, the linearized-map matrix $M$
$(\equiv DP^q)$
has the following block-diagonalized form,
\begin{equation}
M = \left(
   \begin{array} {cc}
   M_1 & {\bf 0} \\
   {\bf 0} & M_2
   \end{array}
   \right),
\label{eq:M}
\end{equation}
where each $2 \times 2$ submatrix $M_i$ $[\equiv W_i(q)]$ $(i=1,2)$ is
calculated through integration of Eq.~(\ref{eq:WEQ}) over the period
$q$. In order to determine the eigenvalues of $M$, it is sufficient to
solve the eigenvalue problems for the two submatrices $M_1$ and $M_2$,
independently. Here the submatrices $M_1$ and $M_2$ determine the
stability of the synchronous orbit against the synchronous-mode and
asynchronous-mode perturbations, respectively. Note also that the
first submatrix $M_1$ is
just the linearized Poincar{\' e} map of the PFDP \cite{Kim1}, and
the coupling affects only the second submatrix $M_2$.

The characteristic equation of each submatrix $M_i$ $(i=1,2)$ is
\begin{equation}
\lambda_i^2 - {\rm tr}M_i \, \lambda_i + {\rm det} \, M_i = 0,
\end{equation}
where ${\rm tr}M_i$ and ${\rm det}M_i$ denote the trace and
determinant of $M_i$, respectively.
As shown in \cite{Lefschetz2}, ${\rm det}\,M_i$ is calculated from a
formula
\begin{equation}
{\rm det}\,M_i = e^{\int_0^q {\rm tr}\,J_i dt}.
\label{eq:Det}
\end{equation}
Substituting the trace of $J_i$ (i.e., ${\rm tr} J_i = - 2 \pi
\gamma$) into
Eq.~(\ref{eq:Det}), we obtain
\begin{equation}
{\rm det}\,M_1 = {\rm det}\,M_2 = e^{-2 \pi \gamma q}.
\label{eq:det}
\end{equation}
Hence, both the submatrices $M_1$ and $M_2$ have the same constant
Jacobian determinant (less than unity).
The eigenvalues, $\lambda_{i,1}$ and $\lambda_{i,2}$, of $M_i$ are
called the Floquet stability multipliers of the synchronous
$q$-periodic orbit.
The first (second) pair of stability multipliers
$(\lambda_{1,1},\lambda_{1,2})$ [$(\lambda_{2,1},\lambda_{2,2})$] of
$M_1$ $(M_2)$ is associated with stability against the
synchronous-mode (asynchronous-mode) perturbation, and hence it may
be called the pair of
synchronous (asynchronous) stability multipliers. Note also that the
pair of synchronous stability multipliers is just the pair of
stability multipliers
of the uncoupled PFDP \cite{Kim1}, and the coupling affects only the
pair of asynchronous stability multipliers.

Each pair of stability multipliers $(\lambda_{i,1},\lambda_{i,2})$
$(i=1,2)$ lies either on the circle of radius $e^{-\pi \gamma q}$, or
on the real axis in the complex plane. The synchronous orbit is stable
when it is stable
against both the synchronous-mode and asynchronous-mode perturbations,
i.e., the moduli of all its four stability multipliers are less than
unity. We first note that all the stability multipliers never cross
the unit circle in
the complex plane, and hence Hopf bifurcations do not occur.
Consequently, the synchronous orbit can lose its stability only when
a multiplier decreases (increases) through $-1$ $(1)$ on the real
axis.

Associate with each pair of stability multipliers
$(\lambda_{i,1},\lambda_{i,2})$ a quantity $R_i$, called the residue,
\begin{equation}
R_i \equiv { {1 + {\rm det}M_i - {\rm tr}M_i} \over
{2(1+{\rm det}M_i)}},
\label{eq:R}
\end{equation}
which was introduced in \cite{Kim3} to characterize stability of
periodic orbits in 2D dissipative maps with constant Jacobian
determinants.
Here the first and second residues $R_1$ and $R_2$ are associated with
stability of the synchronous orbit against the synchronous-mode and
asynchronous-mode perturbations, respectively. Hereafter, they will be
called the synchronous and asynchronous residues, respectively.
Note also that the synchronous residue $R_1$ is just the residue of
the uncoupled PFDP \cite{Kim1}, and the coupling affects only the
asynchronous residue $R_2$. A synchronous periodic orbit is stable
when $ 0 < R_i <1 $ for $i=1,2$; at both ends of $R_i=0$ and $1$, the
stability multipliers $\lambda_i$'s are $1$ and $-1$, respectively.
When each residue $R_i$ decreases through $0$ (i.e., $\lambda_i$
increases through $1$), the periodic
orbit loses its stability via saddle-noddle or pitchfork or
transcritical bifurcation. On the other hand, when $R_i$ increases
through $1$ (i.e.,
$\lambda_i$ decreases through $-1$), it becomes unstable via PDB, also
referred to as a flip or subharmonic bifurcation. For each case of the
PDB's and the pitchfork bifurcations (PFB's), two types of
supercritical and subcritical bifurcations occur. (For more details on
bifurcations in 2D
dissipative maps, refer to Ref.~\cite{Gukenheimer}.)

The stable region of a synchronous periodic orbit in the $A-c$ plane
is bounded by bifurcation lines associated with PDB's and PFB's
(i.e., those curves determined by the equations $R_i=0$ and
$1$ for $i=1,2$), as will be seen in Sec.~\ref{sec:CBTC}.
When the boundary lines on which $R_1$ $(R_2)=0$ and $1$ are crossed,
the synchronous orbit loses its stability via synchronous
(asynchronous) PFB and PDB, respectively. For each case of the
synchronous (asynchronous) PFB
and PDB, two types of supercritical and subcritical bifurcations take
place. In the supercritical case of the synchronous (asynchronous)
PFB and PDB, the synchronous orbit loses its stability, and gives rise
to the birth of a pair of new stable synchronous (asynchronous) orbits
with the same period and a new stable synchronous (asynchronous)
period-doubled orbit,
respectively. However, in the subcritical case of the synchronous
(asynchronous) PFB and PDB, the synchronous orbit becomes unstable by
absorbing a pair of unstable synchronous (asynchronous) orbits with
the same period and an unstable synchronous (asynchronous)
period-doubled orbit, respectively.

Finally, we briefly discuss Lyapunov exponents of a synchronous orbit
in the Poincar{\'e} map $P$, characterizing the mean exponential rate
of divergence of nearby orbits \cite{Lyapunov}. The synchronous and
asynchronous modes of
a nearby orbit are decoupled, because the linearized Poincar{\' e} map
$DP$ at the synchronous orbit is just the block-diagonalized matirx $M$
of Eq.~(\ref{eq:M}) with $q=1$. Therefore, the $2 \times 2$ submatrices
$M_1$ and $M_2$ of $M$ determine the pairs of synchronous and
asynchronous Lyapunov exponents $(\sigma_{1,1},\sigma_{1,2})$ and
$(\sigma_{2,1},\sigma_{2,2})$, characterizing the average exponential
rate of divergence of the synchronous and asynchronous modes of a
nearby orbit, respectively, where $\sigma_{i,1} \geq \sigma_{i,2}$ for
$i=1,2$. Since the two submatrices have the same constant Jacobian
determinant of Eq.~(\ref{eq:det}), each pair of the Lyapunov exponents
satisfy $\sigma_{i,1} + \sigma_{i,2} = - 2 \pi \gamma$ $(i=1,2)$.
Note also that the first pair of synchronous Lyapunov exponents
$(\sigma_{1,1},\sigma_{1,2})$ is just the pair of the Lyapunov
exponents of the uncoupled PFDP, and the coupling affects only the
second pair of
asynchronous Lyapunov exponents $(\sigma_{2,1},\sigma_{2,2})$.

\section{Critical behaviors in two coupled PFDP's}
\label{sec:CBTC}

In this section, by varying the two parameters $A$ and $c$, we study
the critical behaviors of the synchronous supercritical PDB's in the
two coupled PFDP's (\ref{eq:TPFDP1}) for
$\gamma=0.1$ and $\omega_0 = 0.5$. The two coupled PFDP's exhibit
multiple period-doubling transitions to chaos, which is in contrast
to the case of the coupled 1D maps with only single period-doubling
transition to chaos \cite{KK,Kim2}.
For each transition to chaos, the
zero-coupling critical point and an infinity of critical line segments
constitute the critical set in the $A-c$ plane. There exist three
kinds of critical behaviors, depending on the position on the critical
set. These critical behaviors are found to be the same as those of the
two coupled 1D maps \cite{Kim2}.

We consider a linearly-coupled case in which the coupling function
(\ref{eq:CFTN}) is
\begin{equation}
g(x_1,x_2) =  {c \over 2} (x_2 - x_1).
\label{eq:LC}
\end{equation}
Figure \ref{SD1} shows the stability diagram of the synchronous orbits
with low period $q=1,2$. The stable region of a synchronous orbit is
bounded by its PDB and PFB lines. The horizontal (non-horizontal)
solid and short-dashed boundary lines correspond to synchronous
(asynchronous) PDB and PFB lines, respectively. (Each bifurcation may
be supercritical or
subcritical.) Note also that the horizontal synchronous PDB or PFB
lines extend to the (plus) infinity $(c=\infty)$. For the sake of
convenience, only some parts (up to $c=7$) of the infinitely long
lines are drawn in the figure.

We first consider the bifurcations associated with stability of the
stationary point $[(x_1,y_1,x_2,y_2) = (0,0,0,0)]$. Its stable region
is denoted by SP in Fig.~\ref{SD1}. When the non-horizontal
short-dashed line of the
SP is crossed, the stationary point becomes unstable via asynchronous
subcritical PFB. However, at the horizontal solid
boundary line, it loses its stability via synchronous
supercritical PDB, and a new stable synchronous orbit of period $2$
appears. The $2$-periodic orbit is a symmetric orbit with respect to
the inversion symmetry $S_2$, and its stable region is denoted by SP2
in Fig.~\ref{SD1}. When the horizontal (non-horizontal) short-dashed
boundary line of the SP2
is crossed, the symmetric 2-periodic orbit loses its stability via
synchronous (asynchronous) supercritical PFB, and a pair of new stable
synchronous (asynchronous) orbits with period $2$ appears.
Note that the new pair of synchronous orbits is a conjugate pair of
asymmteric orbits with respect to the $S_2$-symmetry.
Its stable region is denoted by ASP2 in Fig.~\ref{SD1}. Unlike the
cases of the
lower-level stability regions (SP and SP2), the ASP2 is $U$-shaped,
because a parabolalike asynchronous PDB line also is a boundary line
of the ASP2. An asynchronous supercritical PDB occurs at the
parabolalike solid line, whereas an asynchronous subcritical PFB
takes place at the non-horizontal short-dashed line. However, each
synchronous asymmetric $2$-periodic orbit becomes unstable via
synchronous supercritical PDB when the horizontal solid line is
crossed, and gives rise
to the birth of a new synchronous asymmetric $4$-periodic orbit.
Here we are interested in such synchronous supercritical PDB's.

Figure \ref{SD2} shows the stability diagram of synchronous asymmetric
orbits born by synchronous supercritical PDB's.
Each synchronous asymmetric orbit of level $n$
(period $2^n$, $n=1,2,3,\dots$)
loses its stability at the horizontal solid line of its stable region
via synchronous supercritical PDB, and gives rise to the birth of a
synchronous asymmetric period-doubled orbit of level $n+1$. Such an
infinite sequence ends
at a finite value of $A^*_1 = 0.357\,709\,845\,3$, which is the first
period-doubling transition point of the uncoupled PFDP \cite{Kim1}.
Consequently, a synchronous quasiperiodic orbit, whose maximum
synchronous
Lyapunov exponent is zero (i.e., $\sigma_{1,1}=0$), exists on the
$A=A^*_1$ line.

We examine the treelike structure of the stability diagram in
Fig.~\ref{SD2}, which consists of an infinite pile of $U$-shape
regions and rectangular-shape
regions. Note that the treelike structure is asymptotically the same
as that in the coupled 1D maps \cite{Kim2}.
The $U$-shape branching is repeated at one side of each $U$-shape
region, including the $c=0$ line segment. The branching side will be
referred to as the zero $c$ side. However, the other side of each
$U$-shape region grows like a chimney without any further branchings
[as an example, see the branch starting from the right side of the
ASP2 in Fig.~\ref{SD2}(b)]. As in the coupled 1D maps \cite{Kim2},
this rule governs the asymptotic
behavior of the treelike structure, even though there are a few
exceptions for lower-level orbits.
Other type of $U$-shape regions without the zero $c$ sides [e.g., the
leftmost $U$-shape region in the third-level stability region in
Fig.~\ref{SD2}(a)] may appear
in the lower-level stability regions. However, the $U$-shape branching
for this kind of $U$-shape region ends at some finite level, and then
each side of the $U$-shape region grows like a chimney without any
further branchings. Consequently, an infinite number of successive
branchings occur only for the case of the $U$-shape region with the
zero $c$ side.

A sequence of connected stability regions with increasing period is
called a ``period-doubling route'' \cite{Kim2}. There are two kinds of
period-doubling routes. The sequence of the $U$-shape regions with the
zero $c$ sides converges to the zero-coupling point $c=0$ on the
$A=A^*_1$ line. It will be referred to as the $U$ route. On the other
hand, a sequence of rectangular regions in each chimney converges to a
critical line segment on the $A=A^*_1$ line. For examples, the
rightmost one in
Fig.~\ref{SD2}(a) is the line segment joining the left end point
$c_l$ $(=0.343\,687 \cdots)$ and the right end point
$c_r$ $(=0.484\,777 \cdots)$ on the $A=A^*_1$ line, and the one in
Fig.~\ref{SD2}(b) is the infinitely-long line connecting the two end
points $c_l$ $(=4.407\,457 \cdots)$ and $c_r$ $(=\infty)$ on the
$A=A^*_1$ line. This kind of route will be called a $C$ route. Note
that there are infinitely many $C$ routes, while the $U$ route
converging to the zero-coupling critical point $(A^*_1,0)$ is unique.
Hence, an infinite number of critical line segments, together with the
zero-coupling critical point, constitute the critical set.

We now study the critical behaviors on the critical set.
First, consider the case of the $U$ route ending at the zero-coupling
critical point. We follow the synchronous orbits of period $q=2^n$ up
to level $n=9$ in the $U$ route, and obtain a self-similar sequence of
parameters $(A_n,c_n)$, at which each orbit of level $n$ has some
given residues $R_1$ and $R_2$ (e.g., $R_1=1$ and $R_2=0$).
Then the sequence $\{ (A_n,c_n) \}$ converges geometrically to the
zero-coupling critical point $(A^*_1,0)$. In order to see the
convergence of each of the two scalar sequences $\{ A_n \}$ and
$\{ c_n \}$, we define
\begin{equation}
\delta_n \equiv { {\Delta A_{n}} \over {\Delta
A_{n+1}} },\;\;\mu_n \equiv { {\Delta c_{n}} \over {\Delta c_{n+1}} },
\end{equation}
where $\Delta A_n = A_n - A_{n-1}$ and $\Delta c_n = c_n - c_{n-1}$.
The sequences of $\delta_n$ and $\mu_n$ are listed in
Table \ref{table1}, and converge
to their limit values, $\delta$ $(\simeq 4.67)$ and $\mu$
$(\simeq -2.5)$, respectively. Hence the two sequences $\{ A_n \}$
and $\{ c_n \}$ obey one-term scaling laws asymptotically:
\begin{equation}
 \Delta A_n \sim \delta^{-n}, \;\; \Delta c_n \sim \mu^{-n}\;\;
 {\rm for \; large\;} n.
 \label{eq:OTSL}
 \end{equation}
As in the coupled 1D maps, the value of the nonlinearity-parameter
scaling factor $\delta$ agrees well with the Feigenbaum constant
$(=4.669 \cdots)$ of the 1D maps \cite{Feigenbaum}.
The value of the coupling-parameter scaling factor $\mu$
is also close to that $(=-2.502\cdots)$ of the coupling-parameter
scaling factor $\alpha$ of the coupled 1D maps near the zero-coupling
critical point \cite{Kim2}. It has been also shown in \cite{Kim2} that
the scaling factor $\alpha$ is just the first relevant ``coupling
eigenvalue'' (CE) $\nu_1$ of the zero-coupling fixed map of the
renormalization transformation for the case of the coupled 1D maps. In
addition to $\nu_1=\alpha$, the
zero-coupling fixed map has another second relevant CE $\nu_2$ $(=2)$,
which also affects the scaling associated with coupling in the coupled
1D map \cite{Kim4}.

In order to get a correction to the leading scaling (\ref{eq:OTSL}),
we take into account the effect of the second relevant CE
$\nu_2$ $(=2)$ on the scaling of the sequence $\{ \Delta c_n \}$ and
extend the simple one-term scaling law (\ref{eq:OTSL}) to a two-term
scaling law \cite{Kim4,Kim5}:
\begin{equation}
\Delta c_n \sim C_1 \mu_{1}^{-n} + C_2 \mu_{2}^{-n} \;\;\;{\rm for
\;large\;}n,
\label{eq:TTSL1}
\end{equation}
where $| \mu_2 | > | \mu_1 | $, and $C_1$ and $C_2$ are some
constants. This is a kind of multiple scaling law \cite{Mao}.
Eq.~(\ref{eq:TTSL1}) gives
\begin{equation}
\Delta c_n = s_1 \Delta c_{n+1} -s_2 \Delta c_{n+2},
\label{eq:RE}
\end{equation}
where $s_1 = \mu_1 + \mu_2$ and $s_2 = \mu_1 \mu_2$.
Then, $\mu_1$ and $\mu_2$ are solutions of the following quadratic
equation,
\begin{equation}
\mu^2 - s_1 \mu + s_2 =0.
\label{eq:QE}
\end{equation}
To evaluate $\mu_1$ and $\mu_2$, we first obtain $s_1$ and $s_2$ from
$\Delta c_n$'s using Eq.~(\ref{eq:RE}):
\begin{equation}
s_1 = { {\Delta c_n \Delta c_{n+1} - \Delta c_{n-1} \Delta c_{n+2}}
\over {\Delta c_{n+1}^2 - \Delta c_n \Delta c_{n+2}} }, \;\;
s_2 = { {\Delta c_n^2 - \Delta c_{n+1} \Delta c_{n-1}}
\over {\Delta c_{n+1}^2 - \Delta c_n \Delta c_{n+2}} }.
\label{eq:T1T2}
\end{equation}
Note that Eqs.~(\ref{eq:TTSL1})-(\ref{eq:T1T2}) hold only for large
$n$. In fact, the values of $s_i$'s and $\mu_i$'s $(i=1,2)$ depend
on the level $n$. Therefore, we explicitly denote $s_i$'s and
$\mu_i$'s by $s_{i,n}$'s and $\mu_{i,n}$'s, respectively. Then, each
of them converges to a constant as $n \rightarrow \infty$:
\begin{equation}
\lim_{n \rightarrow \infty} s_{i,n} = s_i, \;\;\;
\lim_{n \rightarrow \infty} \mu_{i,n} = \mu_i,\;\;i=1,2.
\end{equation}

Three sequences $\{ \mu_{1,n} \}$, $\{ \mu_{2,n} \}$, and
$\displaystyle{ \{ {\mu_{1,n}^2 / \mu_{2,n}} \} }$ are shown in
Table \ref{table2}.
The second column shows rapid convergence of the first scaling factor
$\mu_{1,n}$ to its limit value $\mu_1$ $(\simeq -2.50)$, which agrees
well with the first relevant CE $\nu_1$ ($=\alpha$). (Its convergence
to $\alpha$ is faster than that for the case of the above one-term
scaling law.) The second scaling factor $\mu_{2,n}$ also seems to
converge slowly to its
limit value $\mu$ $(\simeq 3.1)$, whose accuracy is lower than that of
$\mu_1$. As in the coupled area-preserving maps \cite{Kim5}, it seems
from the third and fourth columns that the second scaling factor
$\mu_2$ may be expressed by a product of two relevant CE's
$\nu_1$ $(=\alpha)$ and $\nu_2$ $(=2)$,
\begin{equation}
\mu_2 = {\nu_1^2 \over \nu_2}.
\end{equation}
It has been known that every scaling factor in the multiple-scaling
expansion of a parameter is expressed by a product of the eigenvalues
of a linearized renormalization operator \cite{Mao}.

We also study the coupling effect on the asynchronous residue
$R_{2,n}$ of the synchronous orbit of period $2^n$ near the
zero-coupling critical point
$(A^*_1 , 0)$. Figure \ref{AR1} shows three plots of
$R_{2,n} (A^*_1,c)$ versus
$c$ for $n=4,5,$ and $6$. For $c=0$, $R_{2,n}$ converges to a constant
$R_2^*$ $(=1.300\,59\dots)$, called the critical asynchronous residue,
as $n \rightarrow \infty$. However, when $c$ is nonzero $R_{2,n}$
diverges as $n \rightarrow \infty$, i.e., its slope $S_n$
$\displaystyle{ \left.
(\equiv  {{\partial R_{2,n}} \over {\partial c}}
\right|_{(a^* ,0)})
}$ at the zero-coupling critical point diverges as
$n \rightarrow \infty$.

As in the coupled area-preserving maps \cite{Kim5}, the sequence
$\{ S_n \}$ also obeys a two-term scaling law,
\begin{equation}
S_n = D_1 \nu_1 ^n + D_2 \nu_2 ^n\;\;\;{\rm for\;large\;}n,
\label{eq:TTSL2}
\end{equation}
where $|\nu_1| > |\nu_2|$.
This equation gives
\begin{equation}
S_{n+2} = r_1 S_{n+1} - r_2 S_{n},
\end{equation}
where $r_1 = \nu_1 + \nu_2$ and $r_2 = \nu_1 \nu_2$.
As in the scaling for the coupling parameter, we first obtain
$r_1$ and $r_2$ of level $n$ from $S_n$'s:
\begin{equation}
r_{1,n} = {  {S_{n+1} S_{n} - S_{n+2} S_{n-1}}
\over {S_{n}^2 - S_{n+1} S_{n-1}} },\;\;\;
r_{2,n} = { { S_{n+1}^2 - S_{n} S_{n+2}}
\over {S_{n}^2 - S_{n+1} S_{n-1}} }.
\end{equation}
Then, the scaling factors $\nu_{1,n}$ and $\nu_{2,n}$ of level $n$ are
given by the roots of the quadratic equation,
\begin{equation}
\nu_n^2 - {r_{1,n}} {\nu_n} + {r_{2,n}} =0.
\end{equation}
They are listed in Table \ref{table3} and converge to
constants $\nu_1$ $(\simeq -2.503)$ and $\nu_2$ $(\simeq 2)$
as $n \rightarrow \infty$, whose accuracies are higher than those of
the coupling-parameter scaling factors. Note that the values of
$\nu_1$ and $\nu_2$ agree well with those of the two relevant CE's
$\nu_1$ and $\nu_2$.

We next consider the cases of $C$ routes, each of which converges to a
critical line segment. Two kinds of additional critical behaviors are
found at each critical line segment; the one critical behavior exists
at both ends and the
other critical behavior exists at interior points. In each $C$ route,
there are two kinds
of self-similar sequences of parameters $(A_n,c_n)$, at which each
synchronous orbit of level $n$ has some given residues $R_1$ and
$R_2$; the one converges to the left end point of the critical line
segment and the other converges to
the right end point. As an example, consider the rightmost $C$ route
in Fig.~\ref{SD2}(a), which converges to the critical line segment
with two ends $(A^*_1,c_l)$ and $(A^*_1,c_r)$.
We follow, in the rightmost $C$ route, two self-similar sequences of
parameters, one converging to the left end and the other converging
to the right end. In both cases, the sequence $\{ A_n \}$ converges
geometrically to its accumulation value $A^*_1$ with the 1D scaling
factor $\delta$ $(\simeq 4.67)$ like the case of the $U$ route,
\begin{equation}
\Delta A_n \sim \delta^{-n} \;\;{\rm for\;large}\; n,
\end{equation}
where $\Delta A_n = A_n - A_{n-1}$.
The sequences $\{ c_n \}$ for both cases also obey the one-term
scaling law,
\begin{equation}
\Delta c_n \sim \mu^{-n} \;\;{\rm for\;large}\; n,
\end{equation}
where $\Delta c_n = c_n - c_{n-1}$.
The sequence of the scaling factor $\mu_n$ of level $n$ is listed in
Table \ref{table4}, and converges to its limit value
$\mu$ $(\simeq 2)$. Since the value of the coupling-parameter scaling
factor $\mu$ is different
from that $(\mu=\alpha)$ for the zero-coupling case, the critical
behavior at both ends
differs from that at the zero-coupling critical point.
We also note that the value of $\mu$ agrees well with that of the
coupling-parameter scaling factor ($\nu=2$) of the coupled 1D maps
near both ends of each critical line segment \cite{Kim2}. It has been
also shown in \cite{Kim2} that the scaling factor $\nu$ $(=2)$ is just
the only relevant CE of a nonzero-coupling fixed map of the
renormalization transformation for the case of the coupled 1D maps.

Figure \ref{AR2} shows the behavior of the asynchronous residue
$R_{2,n}(A^*_1,c)$ of the synchronous orbit of period $2^n$ near the
rightmost critical line segment in Fig.~\ref{SD2}(a). For $c=c_l$ and
$c_r$, $R_{2,n}$ converges to a critical residue $R_2^*$
$(=0)$ as $n \rightarrow \infty$,
which is different from that for the zero-coupling case.
The slopes $S_n$'s of $R_{2,n}$ at both ends obey well the one-term
scaling law,
\begin{equation}
S_n \sim \nu^n \;\;{\rm for\; large\;} n.
\end{equation}
The two sequences of the scaling factors $\nu_n$ of level $n$ at both
ends are listed in Table V, and converge to their limit values
$\nu \simeq 2$, which agrees well with that of the only CE ($\nu=2$)
of the nonzero-coupling fixed map governing the critical behavior at
both ends for the case of the
1D maps. However, for any fixed value of $c$ inside the critical line
segment, $R_{2,n}$ converges to a critical residue $R_2^*$ $(=0.5)$ as
$n \rightarrow \infty$ (see Fig.~\ref{AR2}). This superstable case of
$R^*_2=0.5$
corresponds to the supercritical case of $\lambda^*_2=0$
$(\lambda^*_2$: the
second critical stability multiplier) for the coupled 1D maps
\cite{Kim2},
because Eq.~(\ref{eq:R}) of $R$ for the case of 2D maps reduces to the
equation of $R=0.5*(1-\lambda)$ for the case of 1D maps. We also note
that as in the case of the coupled 1D maps, there exists no scaling
factor of the coupling parameter inside the critical line segemnt, and
hence the coupling
parameter becomes an irrelevant one at interior critical points. Thus,
the critical behavior inside the critical line segment becomes the
same as that of the uncoupled
PFDP (i.e., that of the 1D map), which will be discussed in more
details below. This kind of 1D-like critical behavior was found to be
governed by another nonzero-coupling fixed map with no relevant CE for
the case of the coupled 1D maps \cite{Kim2}.

There exists a synchronous quasiperiodic orbit on the $A=A^*_1$ line.
As mentioned in Sec.~\ref{sec:TC}, its synchronous Lyapunov exponents
are the same as the Lyapunov exponents of the uncoupled PFDP, i.e.,
$\sigma_{1,1}=0$ and $\sigma_{1,2}=-0.2 \pi$. The coupling affects
only the second pair of asynchronous Lyapunov exponents
$(\sigma_{2,1},\sigma_{2,2})$,
characterizing the mean exponential rate of divergence of the
asynchronous mode of a nearby orbit. The maximum asynchronous Lyapunov
exponent $\sigma_{2,1}$ near the rightmost critical line segment in
Fig.~\ref{SD2}(a) is shown
in Fig.~\ref{ALYP1}. Inside the critical line segment
$(c_l < c < c_r)$, the
synchronous quasiperiodic orbit on the synchronous plane becomes a
synchronous attractor with $\sigma_{2,1}<0$. Since the dynamics on the
synchronous attractor is the same as that of the uncoupled PFDP, the
critical maps at interior points exhibit essentially 1D-like critical
behaviors, because
the critical behavior of the uncoupled PFDP is the same as that of the
1D maps \cite{Kim1}.
However, as the coupling parameter $c$ passes through $c_l$ and $c_r$,
the maximum asynchronous Lyapunov exponent $\sigma_{2,1}$ of the
synchronous quasiperiodic orbit increases from zero. Hence, the
synchronous quasiperiodic orbit becomes unstable and ceases to be an
attractor outside the critical
line segment. Consequently, the system of the two coupled PFDP's is
asymptotically attracted to another synchronous or asynchronous
attractor outside the critical line. For example, the asymptotic state
for $c=0.343\,68$ $(<c_l)$ becomes an asynchronous attractor of period
512, while that for $c=0.484\,79$ $(>c_r)$ becomes a synchronous
rotational attractor of period 1.

We also study the critical scaling behaviors of the maximum
asynchronous Lyapunov exponent $\sigma_{2,1}$ near both ends of the
rightmost critical line segment in Fig.~\ref{SD2}(a). As shown in
Fig.~\ref{ALYP2},
$\sigma_{2,1}$ varies linearly with respect to $c$ near both ends,
i.e., $\sigma_{2,1} \sim \epsilon$, $\epsilon \equiv c - c^*$
$(c^*=c_l$ or $c_r$).
The critical exponent of $\sigma_{2,1}$ near both ends can be also
obtained from the only CE $\nu=2$ of the nonzero-coupling fixed map
governing the critical behavior near both ends for the case of the
coupled 1D maps. Consider a system with nonzero $\epsilon$ (but with
$A=A^*_1$) near both ends. It is then transformed into a new one of
the same form, but with a renormalized parameter $\epsilon'$ under a
renormalization transformation.
Here the parameter $\epsilon$ obeys a scaling law,
\begin{equation}
\epsilon' = \nu \epsilon = 2\, \epsilon.
\end{equation}
Then the maximum asynchronous Lyapunov exponent $\sigma_{2,1}$
satisfies the homogeneity relation,
\begin{equation}
\sigma_{2,1}({\epsilon}')=2\, \sigma_{2,1}(\epsilon).
\end{equation}
This leads to the scaling relation,
\begin{equation}
\sigma_{2,1} \sim \epsilon^{\eta},
\end{equation}
with critical exponent
\begin{equation}
 \eta = \ln 2 / \ln \nu =1.
\end{equation}

As the nonlinearity parameter $A$ is further increased from
$A=A^*_1$, the stationary point $[x_1=x_2=x^*=0,$ $y_1=y_2=y^*=0$]
undergoes a cascade of
``resurrections'', i.e., it will restabilize after it loses its
stability, destabilize again, and so {\it ad infinitum} \cite{Kim1}.
It was found in \cite{Kim1} that for $\omega_0=0.5$, its
restabilizations occur through
alternating synchronous subcritical PDB's and PFB's, while the
destabilizations take place via alternating synchronous supercritical
PDB's and PFB's. Consequently, the two coupled PFDP's exhibit multiple
period-doubling transitions to chaos. This is in contrast to the case
of the coupled 1D maps, in which only single period-doubling
transition to chaos occurs \cite{KK,Kim2}.

As the first example of the multiple period-doubling transitions to
chaos, we consider the first resurrection of the
stationary point shown in Fig.~\ref{BDSD1}(a). For $A=A_r(1)$
$(=3.150\,509 \cdots)$, a
synchronous subcritical PDB occurs. Hence, the stationary point
restabilizes with birth of a new unstable synchronous symmetric orbit
of period $2$ for $A>A_r(1)$. As $A$ is increased from $A=A_r(1)$, the
stationary point destabilizes at $A=A_d(2)$ $(=3.224\,230 \cdots)$ via
synchronous supercritical PFB, which results in the birth of a
conjugate pair of synchronous asymmetric orbits with period $1$.
Fig.~\ref{BDSD1}(b) shows the stability diagram of the stationary point
and the synchronous asymmetric orbits of level $n$ (period $2^n$,
$n=0,1,2,3,4$) near the $c=0$ line in the $A-c$ plane \cite{Rem}.
Each synchronous asymmetric orbit of level $n$ becomes unstable at the
horizontal solid line of its stable region via synchronous
supercritical PDB,
and gives rise to the birth of a synchronous asymmetric period-doubled
orbit of level $n+1$. Such an infinite sequence terminates at a finite
value of $A^*_2 = 3.263\,703\,15 \cdots$, which is the second
period-doubling transition point of the uncoupled PFDP \cite{Kim1}.
Note that the treelike structure of the stability diagram in
Fig.~\ref{BDSD1}(b) is
essentially the same as that in Fig.~\ref{SD2}(a). Hence, the critical
set also
consists of an infinite number of critical line segments and the
zero-coupling critical point, as in the first period-doubling
transition case. In order to study the critical behaviors on the
critical set, we follow the synchronous
asymmetric orbits up to level $n=7$ in the $U$ route and the rightmost
$C$ route. It is found that the critical behaviors are the same as
those for the first period-doubling transition case. That is, there
exist three kinds of critical behaviors at the zero-coupling critical
point, both ends of each critical line
segment and interior points.

As the second example, we also consider the second resurrection of the
stationary point shown in Fig.~\ref{BDSD2}(a). A synchronous
subcritical PFB takes
place at $A=A_r(2)$ $(=10.093\,985 \cdots)$. Consequently, the
stationary point restabilizes with birth of a pair of new unstable
orbits with period
$1$. As $A$ is further increased, the stationary point destabilizes at
$A=A_d(3)$ $(=10.097\,583 \cdots)$ via synchronous supercritical PDB,
which results in the birth of a new synchronous symmetric orbit with
period $2$. The subsequent bifurcation behaviors are the same as those
for the first period-doubling transition case. That is, a third
infinite sequence of
synchronous supercritical PDB's follows and ends at a finite value
$A^*_3$ $(=10.099\,660\,93 \cdots)$, which is the third
period-doubling transition point of the uncoupled PFDP \cite{Kim1}.
The third stability diagram of synchronous orbits near the $c=0$ line
is shown in Fig.~\ref{BDSD2}(b) \cite{Rem}. Note
that its treelike structure is essentially the same as that in
Fig.~\ref{SD2}(a). Hence, the critical
set is composed of the zero-coupling critical point and an infinity of
critical line segments. Furthermore, the critical behaviors on the
critical set are found
to be the same as those for the first period-doubling transition case.

In addition to the linear-coupling case (\ref{eq:LC}), we have also
studied other nonlinear-coupling cases,
\begin{equation}
g(x_1,x_2) = {c \over 2} [x_2^n-x_1^n], \;\;n=2,3.
\label{eq:NLC}
\end{equation}
For the first period-doubling transition case, the stability diagrams
of synchronous orbits near the $c=0$ line for the cases of the
quadratic and cubic couplings are shown in Fig.~\ref{NCSD}(a)
and \ref{NCSD}(b), respectively. Their treelike structures are
essentially the same as that in Fig.~\ref{SD2}(a). Hence, the
zero-coupling critical
point and an infinite number of critical line segments constitute the
critical set for each nonlinear-coupling case. Moreover, the critical
behaviors for these
nonlinear-coupling cases are also found to be the same as those for
the linear-coupling case.

\section{Extension to many coupled PFDP's}
\label{sec:CBMC}

In this section we study the critical behaviors of the synchronous
PDB's in $N$-coupled
$(N \geq 3)$ PFDP's in which the coupling extends to the $K$th
$[1 \leq K \leq {N \over 2} ({{N-1} \over 2})$ for even (odd) $N$]
neighbor(s) with equal strength. It is found that the critical
behaviors depend on the
coupling range. In the global-coupling case, in which each PFDP is
coupled to all the other ones with equal coupling strength, the
structure of the critical set and the critical behaviors are the
same as those for the two-coupled case,
independently of $N$. However, for any other nonglobal-coupling cases,
the structure of the critical set becomes different from that for the
global-coupling case, because of a significant change in the stability
diagram.

Consider $N$ symmetrically coupled PFDP's with a periodic boundary
condition,
\begin{equation}
{\ddot{x}}_m = f(x_m,{\dot x}_m,t) + g(x_m,x_{m+1},\dots,x_{m-1}),
\;\;m=1,2,\dots,N.
\label{eq:MCPFDP}
\end{equation}
Here the periodic boundary condition imposes $x_m(t) = x_{m+N}(t)$ for
all $m$, the function $f(x,{\dot x},t)$ is given in
Eq.~(\ref{eq:fftn}), and
$g(x_1,\dots,x_N)$ is a coupling function, obeying the condition
\begin{equation}
g(x,\dots,x) = 0 \;{\rm for\;all\;}x.
\label{eq:CC}
\end{equation}

A general form of coupling for odd $N$ $(N \geq 3)$ is given by
\begin{eqnarray}
g(x_1,\dots,x_N) &=& {\frac {c} {2K+1}}
  {\sum_{l=-K}^{K}} [u(x_{1+l}) - u(x_1)], \nonumber \\
&=& c \left[ {\frac {1}{2K+1}} {\sum_{l=-K}^{K}} u(x_{1+l}) - u(x_1)
\right], \nonumber \\
&&K=1,\dots,{\frac {N-1} {2}},
\label{eq:GCF}
\end{eqnarray}
where $c$ is a coupling parameter and $u$ is a function of one
variable. Here the coupling extends to the $K$th neighbors with
equal coupling strength, and the function $g$ satisfies the condition
(\ref{eq:CC}).
The extreme long-range interaction for
$K= {{\frac {N-1} {2}}}$ is
called a global coupling, for  which  the coupling  function $g$
becomes
\begin{eqnarray}
g(x_1,\dots,x_N)  &=&  {\frac  {c}  {N}}{\sum_{m=1}^{N}}  [u(x_{m}) -
u(x_1)] \nonumber \\
&=& c  \left[ {\frac  {1}{N}} {\sum_{m=1}^{N}} u(x_m)  - u(x_1)
\right].
\label{eq:GC}
\end{eqnarray}
This  is a  kind  of mean-field  coupling,  in which  each  element is 
coupled to all the  other elements  with equal  coupling strength. All
the other couplings with $K < {{\frac {N-1} {2}}}$ (e.g.,
nearest-neighbor
coupling with $K=1$) will be referred to as non-global couplings.
The $K=1$ case for $N=3$ corresponds to both the global coupling
and the nearest-neighbor coupling.

We next consider the case of even $N$ $(N \geq 2)$.
The  form of  coupling of  Eq.~(\ref{eq:GCF}) holds  for the  cases of 
non-global couplings with $K=1,\dots,{{\frac{N-2}{2}}}$ $(N \geq 4)$.
The global coupling for $K= {{\frac {N} {2}}}$ $(N \geq 2)$ also
has the form of Eq.~(\ref{eq:GC}), but it cannot  have the form of
Eq.~(\ref{eq:GCF}), because there exists only one farthest neighbor
for $K= { {\frac{N}{2}} }$, unlike the case of odd $N$.
The  $K=1$ case  for $N=2$  also  corresponds to  the nearest-neighbor 
coupling as well as to the global coupling, like the $N=3$ case.

The stability analysis of an orbit in many coupled PFDP's is
conveniently carried out by Fourier-transforming with respect to the
discrete space $\{m\}$ \cite{Kapral}. Consider an orbit
$\{ {x_m}(t)\;  ; \;m=1,\dots,N \}$ of the $N$ coupled PFDP's
(\ref{eq:MCPFDP}). The discrete spatial Fourier transform of the orbit
is:
\begin{eqnarray}
{\cal F}[{x_m(t)}] &\equiv& {\frac{1}{N}} {\sum_{m=1}^{N}}
{e^{-2{\pi}imj/N}} {x_m}(t) = {\xi}_j(t), \nonumber \\
&&\;\;\;\;\;\;\;\;\;\; j=0,1,\dots,N-1.
\label{eq:FT}
\end{eqnarray}
The Fourier transform $\xi_j(t)$ satisfies $\xi_j^*(t) = \xi_{N-j}(t)$
($*$ denotes complex conjugate), and
the wavelength of a mode with index $j$ is ${\frac {N}{j}}$ for
$j \leq {{ {\frac {N} {2}}}}$ and
${\frac {N} {N-j}}$ for $j > {{\frac {N} {2}}}$.

To  determine the  stability of  a synchronous $q$-periodic orbit
[$x_1(t)  = \cdots =x_N(t) \equiv x^{*}(t)$ for all $t$ and
$x^*(t) = x^*(t+q)$],
we consider an infinitesimal perturbation $\{ {\delta}x_m(t) \}$
to the synchronous orbit, i.e.,
$x_m(t)=x^{*}(t)+{\delta}x_m(t)$ for $m=1,\dots,N$.
Linearizing the $N$-coupled PFDP's (\ref{eq:MCPFDP}) at the
synchronous orbit, we obtain:
\begin{eqnarray}
{\delta}{\ddot x_m} &=& { {\partial f(x^*,\dot{x}^*,t)} \over
{\partial x^*}}
{\delta}x_m + { {\partial f(x^*,\dot{x}^*,t)} \over {\partial
\dot{x}^*} }
{\delta}{\dot x}_m \nonumber \\
 &&   +   {\sum_{l=1}^{N}}   {G_l}(x^{*})\;   {\delta}x_{l+m-1},
\label{eq:LE}
\end{eqnarray}
where
\begin{equation}
G_l(x) \equiv
\left. { \frac{\partial g(x_1,\dots,x_N)}{\partial x_l} }
\right |_{x_1=\cdots=x_N=x}.
\label{eq:RCF}
\end{equation}
Hereafter the  functions $G_l$'s  will be called  ``reduced'' coupling 
functions
of $g(x_1,\dots,x_N)$.

Let  ${\delta  {\xi}_j}(t)$  be  the  Fourier  transform  of  {$\delta 
x_m(t)$},
i.e.,
\begin{eqnarray}
\delta \xi_j =
{\cal F}[{\delta x_m(t)}] &=& {\frac{1}{N}} {\sum_{m=1}^{N}}
{e^{-2{\pi}imj/N}} {\delta x_m}, \nonumber \\
&&\;\;\;\;\;\;\;\;\;j=0,1,\dots,N-1.
\end{eqnarray}
Here $\delta \xi_0$ is the synchronous-mode perturbation, and all the
other $\delta \xi_j$'s with nonzero indices $j$ are the
asynchronous-mode
perturbations. Then the Fourier transform of Eq.~(\ref{eq:LE}) becomes:
\begin{eqnarray}
\delta {\ddot{\xi}}_j &=&
{{\partial f(x^*,\dot{x}^*,t)} \over {\partial \dot{x}^*}}
{\delta}{\dot{\xi}}_j +
[{{\partial f(x^*,\dot{x}^*,t)} \over {\partial x^*}}  \nonumber \\
&& + \sum_{l=1}^{N} {G_l}(x^{*}) {e^{2 \pi i(l-1)j/N}}]
{\delta {\xi}_j}, \;\; j=0,1,\dots,N-1.
\label{eq:LM1}
\end{eqnarray}
Note that all the modes $\delta \xi_j$'s become decoupled for the
synchronous orbit.

The equation (\ref{eq:LM1}) can also be put into the following form:
\begin{eqnarray}
\left(
\begin{array}{l}
\delta{\dot{\xi}}_j \\
\delta{\dot{\eta}}_j
\end{array}
\right)
&=& L_j(t)
\left(
\begin{array}{l}
\delta{\xi}_j \\
\delta{\eta}_j
\end{array}
\right), \;\;j=0,1,\dots,N-1,
\label{eq:LM2}
\end{eqnarray}
where
\begin{equation}
L_j(t)=
\left( \begin{array}{cc}
0 & \;\; 1 \\
{{{\partial f(x^*,\dot{x}^*,t)} \over {\partial x^*} } +
{\displaystyle{\sum_{l=1}^{N}}} {G_l}(x^{*}) {e^{2 \pi i(l-1)j/N}}}
&\;\; {{\partial f(x^*,\dot{x}^*,t)} \over {\partial \dot{x}^*} }
    \end{array}
\right).
\label{eq:JML}
\end{equation}
Note that each $L_j$ is a $q$-periodic matrix, i.e.,
$L_j(t) = L_j(t+q)$.
Let $\Phi_j(t)=(\phi^{(1)}_j(t),\phi^{(2)}_j(t))$ be a fundamental
solution matrix with $\Phi_j(0) = I$. Here $\phi^{(1)}_j(t)$ and
$\phi^{(2)}_j(t)$ are two independent solutions expressed in column
vector forms, and $I$ is
the $2 \times 2$ unit matrix. Then a general solution of the
$q$-periodic system has the following form
\begin{eqnarray}
\left(
\begin{array}{l}
\delta{\xi_j} (t) \\
\delta{\eta_j} (t)
\end{array}
\right)
&=& \Phi_j(t)
\left(
\begin{array}{l}
\delta{\xi}_j (0)\\
\delta{\eta}_j (0)
\end{array}
\right), \nonumber \\
&&\;\;\;\;\;\;\;\;j=0,1,\dots,N-1,
\label{eq:LM3}
\end{eqnarray}
Substitution of Eq.~(\ref{eq:LM3}) into Eq.~(\ref{eq:LM2}) leads to an
initial-value problem to determine $\Phi_j(t)$,
\begin{equation}
{\dot \Phi}_j(t) = L_j(t) \Phi_j(t),~\Phi_j(0)=I.
\label{eq:MCWEQ}
\end{equation}
Each $2 \times 2$ matrix $\Psi_j$ $[\equiv \Phi_j(q)]$, which is
obtained through integration of Eq.~(\ref{eq:MCWEQ}) over the period
$q$, determines the stability of the q-periodic synchronous orbit
against the $j$th-mode perturbation.

The characteristic equation of each matrix $\Psi_j$
$(j=0,1,\dots,N-1)$ is
\begin{equation}
\lambda_j^2 - {\rm tr}\, \Psi_j \, \lambda_j + {\rm det} \, \Psi_J =0,
\end{equation}
where ${\rm tr} \Psi_j$ and ${\rm det} \Psi_j$ denote the trace and
determinant of $\Psi_j$, respectively.
As shown in \cite{Lefschetz2}, ${\rm det}\,\Psi_j$ is given by
\begin{equation}
{\rm det}\,\Psi_j = e^{\int_0^q {\rm tr}\,L_j dt}=
e^{-2 \pi \gamma q}.
\label{eq:MCDet}
\end{equation}
Hence, all the matrices $\Psi_j$'s have the same constant Jacobian
determinant (less than unity).
The eigenvalues, $\lambda_{j,1}$ and $\lambda_{j,2}$, of $\Psi_j$ are
called the Floquet stability multipliers, which are associated with
the stability of the synchronous $q$-periodic orbit against the
$j$th-mode perturbation. Since the $j=0$ case corresponds to the
synchronous mode, the first pair
of stability multipliers $(\lambda_{0,1},\lambda_{0,2})$ is called the
pair of synchronous stability multipliers. On the other hand, all the
other pairs of stability multipliers are called the pairs of
asynchronous stability
multiplies, because all the other cases of $j \neq 0$ correspond to
asynchronous modes. Like the two-coupled case [see Eq.~(\ref{eq:R})],
we also associate with a pair of stability multipliers
$\lambda_{j,1}$ and $\lambda_{j,2}$ a residue $R_j$,
\begin{equation}
R_j \equiv { {1 + {\rm det} \Psi_j - {\rm tr}\Psi_j} \over
{2(1+{\rm det}\Psi_j)}},\;\;j=0,1,\dots,N-1.
\label{eq:MCR}
\end{equation}
Here the first one $R_0$ is associated with the stability against the
synchronous-mode perturbation, and hence it may be called the
synchronous residue.
On the other hand, all the other ones $R_j$ $(j \neq 0)$ are called
the asynchronous residues, because they are associated with the
stability against the asynchronous-mode perturbations.

It follows from the condition (\ref{eq:CC}) that the reduced coupling
functions satisfy
\begin{equation}
\sum_{l=1}^{N} G_l(x) =0.
\end{equation}
Hence the matrix (\ref{eq:JML}) for $j=0$ becomes
\begin{equation}
L_0(t)=
\left( \begin{array}{cc}
0 & \;\; 1 \\
{{\partial f(x^*,\dot{x}^*,t)} \over {\partial x^*} }
& \;\; {{\partial f(x^*,\dot{x}^*,t)} \over {\partial \dot{x}^*} }
    \end{array}
\right).
\end{equation}
This is just the linearized Poincar{\'e} map of the uncoupled PFDP
\cite{Kim1}. Hence the synchronous residue $R_0$ becomes the same as
the residue of the uncoupled PFDP, i.e., it depends only on the
nonlinearity parameter $A$. While there is no coupling effect on
$R_0$, the coupling affects all the other asynchronous residues
$R_j$ $(j \neq 0)$.

In case of the global coupling of Eq.~(\ref{eq:GC}),
the reduced coupling functions become:
\begin{equation}
{G_l}(x) = \left \{
 \begin{array}{l}
  (1-N) G(x)\;\;\;\; {\rm for}\;l=1, \\
  \;\;\;\;\;\;G(x)\;\;\;\;\;\;\;\;\;\;{\rm for}\;l \neq 1,
  \end{array}
  \right.
\end{equation}
where $G(x)= {{\frac{c}{N}}} u'(x)$.
Substituting $G_l$'s into the second term of the $(2,1)$ entry of the
matrix $L_j(t)$, we have:
\begin{equation}
{\sum_{l=1}^{N}} G_l(x) e^{2 \pi i(l-1)j/N} =
\left \{ \begin{array}{l}
          \;\;\;\;\;0\;\;\;\;\;\;\;{\rm for}\;\; j=0, \\
          -c\, u'(x)\;\;{\rm for}\;\; j \neq 0.
         \end{array}
\right.
\label{eq:SE2}
\end{equation}
Hence all the asynchronous residues $R_j$ $(j \neq 0)$ become the
same, i.e., $R_1  =  \cdots = R_{N-1}$.
Consequently, like the two-coupled case, there exist only  two
independent residues $R_0$ and $R_1$, the values of which are
also independent of $N$.

We next  consider the non-global  coupling of  the form (\ref{eq:GCF}) 
and define
\begin{equation}
G(x) \equiv {\frac {c} {2K+1}} u'(x),
\end{equation}
where $1 \leq K \leq {{{\frac {N-2} {2}}}}\;
({{\frac {N-3} {2}}})$ for even (odd) $N$ larger than 3.
Then we have
\begin{equation}
{G_l}(x) = \left \{
 \begin{array}{l}
  -2 K G(x)\;\;\; {\rm for}\;l=1, \\
  \;\;\;\;\;G(x)\;\;\;\;\;\; {\rm for}\;  2 \leq l \leq  1+K \;\; {\rm 
or} \\
  \;\;\;\;\;\;\;\;\;\;\;\;\;\;\;\;\;\;\;{\rm  for}\;N+1-K \leq  l \leq 
N, \\
  \;\;\;\;\;\;\;0\;\;\;\;\;\;\;\;\;\;{\rm otherwise.}
  \end{array}
  \right.
\end{equation}
Substituting the reduced coupling functions into
the  matrix $L_j(t)$  of Eq.~(\ref{eq:JML}),  the  second term  of the 
$(2,1)$ entry of $L_j(t)$ becomes:
\begin{equation}
{\sum_{l=1}^{N}} G_l(x) e^{2 \pi i(l-1)j/N}
= - {S_N}(K,j) c\, u'(x),
\label{eq:SE1}
\end{equation}
where
\begin{equation}
{S_N}(K,j) \equiv {4 \over {2K+1}} {\sum_{k=1}^{K}}
\sin^2 {{\pi jk} \over {N}}
= 1- {\frac {\sin(2K+1) {{\frac{\pi j}{N}}}}
{(2K+1) \sin{{\frac{\pi j}{N}}}}}.
\label{eq:SF}
\end{equation}
Hence, unlike the global-coupling case, all the asynchronous residues
vary depending on the coupling range $K$ as well as on  the mode
number  $j$.
Since $S_N(K,j)  =  S_N(K,N-j)$, the residues satisfy
\begin{equation}
R_j = R_{N-j},\;\;j=0,1,\dots,N-1.
\end{equation}
Thus it is sufficient to consider only the case of
$0 \leq j \leq {N \over 2}$ $({{N-1} \over 2})$ for even (odd) $N$.
Comparing the expression in Eq.~(\ref{eq:SE1}) with that in
Eq.~(\ref{eq:SE2}) for $j \neq 0$, one can easily see that
they are the same except for the factor $S_N (K,j)$. Consequently,
making a change of the coupling parameter
${c \rightarrow {c \over {S_N (K,j)}}}$, the residue $R_j$ for the
non-global coupling case  of range $K$  becomes the same as that
for the global-coupling case.

Each pair of stability multipliers $(\lambda_{j,1},\lambda_{j,2})$
$(j=0,1,\dots,N-1)$ lies either on the circle of radius
$e^{-\pi \gamma q}$, or on the real axis in the complex plane.
The synchronous orbit is stable
against the $j$th-mode perturbation when $0 < R_j <1$ (i.e.,
the pair of stability multipliers $(\lambda_{j,1}, \lambda_{j,2})$
lies inside the unit circle in the complex plane).
A PDB (PFB) occurs when the residue $R_j$ increases (decreases)
through $1$ $(0)$ [i.e., a stability multiplier decreases
(increases) through $-1$ $(1)$]. We also note that a(n) synchronous
(asynchronous) bifurcation takes
place for $j=0$ $(j \neq 0)$. For more details on bifurcatios, refer
to Sec.~\ref{sec:TC}.

When the synchronous residue $R_0$ of a synchronous periodic orbit
increases through $1$, the synchronous orbit loses its stability
via synchronous PDB, giving rise to the birth of a new synchronous
period-doubled orbit.
Here we are interested in such synchronous PDB's. Thus, for each mode
with nonzero index $j$ we consider a region in the $A-c$ plane, in
which the synchronous orbit is stable against the perturbations of
both modes with indices $0$ and $j$. This stable region is bounded by
four bifurcation curves determined by the equations $R_0 = 0,\,1$ and
$R_j=0,\,1$, and it will be denoted by $U_N$.

For the case of global coupling, those stable regions coincide,
irrespectively of $N$ and $j$, because all the asynchronous residues
$R_j$'s $(j \neq 0)$ are the same, independently of $N$. The stable
region for this global-coupling case will be denoted by $U_G$. Note
that $U_G$ itself is just the stability region of
the synchronous orbit, irrespectively of $N$,
because the synchronous orbit is stable against the perturbations of
all synchronous and asynchronous modes in the region $U_G$. Thus the
stability diagram of synchronous orbits of period $2^n$
$(n=1,2,3,\dots)$ in the $A-c$ plane
becomes the same as that for the two-coupled case, independently of
$N$. That is, the stable regions of the synchronous orbits form a
``stability tree'' in the parameter plane [see Figs.~2(a) and 2(b)].
Consequently, the zero-coupling critical point and an infinite number
of critical line segments constitute the critical set.
There exists one kind of critical behavior in the $U$ route ending at
the zero-coupling critical point, while two other kinds of critical
behaviors exist in each $C$ route ending
at a critical line segment. The three kinds of critical behaviors are
the same as those for the two-coupled case, independently of $N$.
For more details on the critical behaviors, refer to
Sec.~\ref{sec:CBTC}.

However, the stable region $U_N$ vary depending on the coupling range
$K$ and the mode number $j$ for the nonglobal-coupling cases,
i.e., $U_N=U_N(K,j)$. To find the stability region of a synchronous
orbit in $N$ coupled PFDP's with a given $K$, one may start with the
stability region $U_G$ for the global-coupling case. Rescaling the
coupling parameter $c$ by a scaling factor $1 \over S_N(K,j)$ for
each nonzero $j$, the stable region $U_G$ is transformed into a
stable region $U_N(K,j)$. Then the stability region of the
synchronous orbit is given by the intersection of all such stable
regions $U_N$'s. An important change occurs
in the stability diagram of the synchronous orbits of period $2^n$
$(n=1,2,\dots)$, and consequently the structure of the critical set
becomes different from that for the global-coupling case, as will
be seen below.

As an example, we consider the nearest-neighbor coupling case with
$K=1$ in four linearly-coupled PFDP's, in which the coupling function
is given by
\begin{equation}
g(x_1,x_2,x_3,x_4) = {c \over 3} (x_2+x_4 - 2 x_1).
\end{equation}
Figure \ref{MCSD} shows the stability regions of the synchronous
$2^n$-periodic
$(n=1,2,3,4)$ orbits. Note that the scaling factor
$1 \over {S_4(1,j)}$ has its minimum value $3 \over 4$ at $j=2$.
However, for each synchronous orbit, $U_4(1,2)$ itself cannot be the
stability region, because bifurcation curves of different
modes with nonzero indices intersect one another. We first examine
the structure of the stability diagram in Fig.~\ref{MCSD}(a), starting
from the left side of the stability region of the synchronous orbit of
level $1$ $(n=1)$.
The zero $c$ side of $U_4(1,2)$ including a $c=0$ line segment remains
unchanged, whereas the other side becomes flattened by the bifurcation
curve of the asynchronous mode with $j=1$. Due to the successive
flattening with increasing level $n$, a significant change in the
stability diagram occurs.
Of the infinite number of period-doubling routes for the
global-coupling
case, only the $U$ route ending at the zero-coupling critical point
remains. Thus only the zero-coupling point is left as a critical point
in the parameter plane. However, as shown in Fig.~\ref{MCSD}(b), the
rightmost branch of the stability
diagram, starting from the right side of the stability region of the
synchronous periodic orbit of level $1$, is the same as that for the
global-coupling case except that the coupling parameter $c$ is
rescaled with the maximum scaling factor $1 \over S_4(1,1)$ $(=1.5)$
of the $j=1$ mode.
Hence, the rightmost $C$ route ending at a critical line segment is
also left. Consequently, the critical set for this linear-coupling
case is composed of the zero-coupling critical point and one critical
line segment.

Consider a self-similar sequence of parameters $(A_n,c_n)$, at which
the synchronous orbits of period $2^n$ has some given residues, in
the $U$ route
for the global-coupling case. Rescaling the coupling parameter with
the minimum scaling factor $S_4(1,2)$ $(=0.75)$, the sequence is
transformed into a self-similar one for the $N=4$ case of
nearest-neighbor coupling. Hence, the critical behavior near the
zero-coupling critical point
becomes the same as that for the global-coupling case. As mentioned
above, the rightmost $C$ route in Fig.~\ref{SD2}(b) for the
global-coupling case is also transformed into
the $C$ route in Fig.~\ref{MCSD}(b) for the nearest-neighbor coupling
case by rescaling $c$ with the maximum scaling factor
$S_4(1,1)$ $(=1.5)$. Hence, the critical behaviors at both ends and
interior points of the critical line segment are the same as those
for the global-coupling case.

The results for the nearest-neighbor coupling case with $K=1$ extends
to all the other nonglobal-coupling cases with
$1 < K < {N \over 2} ({{N-1} \over 2})$
for even (odd) $N$. For each nonglobal-coupling case with $K>1$, we
first consider a mode with index $j_{\rm min}$ for which the scaling
factor $1 \over S_N(K,j)$ becomes the smallest one and the stability
region $U_N(K,j_{\rm min})$ including a $c=0$ line segment. Here the
value of $j_{\rm min}$ varies depending on the range $K$. Like the
$K=1$ case, the zero $c$ side of $U_N(K,j_{\rm min})$ including the
$c=0$ line segemnt remains unchanged, whereas the other side becomes
flattened by the bifurcation curves of the other modes with nonzero
indices. Thus the overall shape of
the stability diagram, starting from the left zero $c$ side of the
stability region of the synchronous $2$-periodic orbit, becomes
essentially the same as that for the nearest-neighbor coupling case.
Consequently, only the $U$
route ending at the zero-coupling critical point is left as a
period-doubling route, and the critical behavior near the
zero-coupling critical point is also the same as
that for the global-coupling case. We next consider a mode with index
$j_{\rm max}$ for which the scaling factor $1 \over S_N(K,j)$ becomes
the largest one. Rescaling $c$ with the maximum scaling factor
$1 \over S_N(K,j_{\rm max})$, the rightmost $C$ route in
Fig.~\ref{SD2}(b) for the
global-coupling case is transformed into the $C$ route for the
nonglobal-coupling case, and the critical behaviors at the critical
line segment are also the same as those for the global-coupling case.

\section{Summary}
\label{sec:Sum}
The critical behaviors of PDB's in $N$ coupled PFDP's are investigated
by varying two parameters $A$ and $c$. As $A$ is increased, the
stationary point of the
coupled PFDP's undergoes an infinite series of period-doubling
transitions to chaos. This is in contrast to the case of the coupled
1D maps with only single period-doubling transition to
chaos \cite{KK,Kim2}.
The two-coupled case with $N=2$ has been first studied. For each
period-doubling transition to chaos, the
zero-coupling critical point and an infinity of critical line segments
constitute the critical set in the parameter plane. There are three
kinds of critical behaviors, depending on the position of the critical
set. They are found to be the same as those for the coupled 1D
maps \cite{Kim2}. We also extend the results
of the two-coupled case to many coupled PFDP's, in which the critical
behaviors vary depending on whether or not the coupling is global. In
the global-coupling case, the critical behaviors are the same as those
for the two-coupled case, independently of $N$. However, for any other
nonglobal-coupling cases, the structure of
the critical set becomes different from that for the global-coupling
case, because of an important change in the stability diagram of
$2^n$-periodic orbits $(n=0,1,2,\dots)$.

\acknowledgments
This work was supported by the Exchange Program of the Senior Scientist,
the Korea Science and Engineering Foundation. One of us (S.Y.K.) thanks
Professor R. Fox and Ms. M. Choi for their hospitality during the period
of his visit to the Georgia Institute of Technology.

%
%

\begin{table}
\caption{ In the $U$ route, we followed a sequence of parameters
          $(A_n,c_n)$ at which the pair of residues
          $(R_{1,n},R_{2,n})$
          of the synchronous orbit of period $2^n$ is $(1,0)$. This
          sequence converges to the zero-coupling critical point
          $(A^*_1,0)$. The scaling factors of the nonlinearity and
          coupling parameters $A$ and $c$ are shown in the second and
          third columns, respectively.
        }
\label{table1}
\begin{tabular}{ccc}
$n$ & $\delta_n$ & $\mu_n$ \\
\tableline
2 & 5.286 &   -2.96 \\
3 & 4.692  &  -2.91  \\
4 & 4.665  &   -2.41 \\
5 & 4.666  &   -2.59  \\
6 & 4.667  &   -2.43  \\
7 & 4.670  &   -2.57   \\
8 & 4.665  &   -2.45
\end{tabular}
\end{table}

\begin{table}
\caption{For the case of the $U$ route, the scaling factors
$\mu_{1,n}$ and $\mu_{2,n}$ in the two-term scaling for the coupling
parameter are shown in the second and third columns, respectively.
         A product of them, ${\mu^2_{1,n}} \over {\mu_{2,n}}$, is
         shown in the fourth column.
        }
\label{table2}
\begin{tabular}{cccc}
$n$ & $\mu_{1,n}$ & $\mu_{2,n}$ & ${\mu^2_{1,n}} \over {\mu_{2,n}}$ \\
\tableline
4 & -2.536 & 6.87 & 0.94 \\
5 & -2.500 & 2.84 & 2.20 \\
6 & -2.500 & 2.81 & 2.22 \\
7 & -2.504 & 3.09 & 2.03
\end{tabular}
\end{table}

\begin{table}
\caption{The scaling factors $\nu_{1,n}$ and $\nu_{2,n}$ in the
two-term scaling for the slope $S_n$ of the asynchronous residue
         $R_{2,n}$ at the zero-coupling critical point are shown in
         the second and third columns, respectively.
        }
\label{table3}
\begin{tabular}{ccc}
$n$ & $\nu_{1,n}$ & $\nu_{2,n}$  \\
\tableline
4 & -2.599 & 2.783  \\
5 & -2.511 & 1.923  \\
6 & -2.503 & 2.004   \\
7 & -2.503 & 1.998   \\
8 & -2.503 & 1.999
\end{tabular}
\end{table}

\begin{table}
\caption{We followed, in the rightmost $C$ route in Fig.~2(a), two
         self-similar sequences of parameters $(A_n,c_n)$, at which
         the pair of residues $(R_{1,n}, R_{2,n})$ of the synchronous
         orbit with period
         $2^n$ is $(1,0.1)$. They converge to both ends of the
         critical line segment. The scaling factors of the coupling
         paramter at the left
         and right ends are shown in the second and third columns,
         respectively. In both cases the scaling factors seem to
         converge to the same limit value $\mu \simeq 2$.
        }
\label{table4}
\begin{tabular}{ccc}
$n$ & $\mu_n$ & $\mu_n$  \\
\tableline
5 & 1.05 & 3.12 \\
6 & 1.76 & 2.55 \\
7 & 1.85 & 2.26 \\
8 & 1.94 & 2.12
\end{tabular}
\end{table}

\begin{table}
\caption{The scaling factors $\nu_n$'s in the one-term scaling for the
slopes $S_n$'s of the asynchronous residue $R_{2,n}$ at the left and
right ends of the rightmost critical line segment in Fig.~2(a) are
shown in the second and third columns, respectively.
        }
\label{table5}
\begin{tabular}{ccc}
$n$ & $\nu_n$ & $\nu_n$  \\
\tableline
4 & 2.156 & 1.991 \\
5 & 1.971 & 2.003 \\
6 & 2.006 & 1.999 \\
7 & 1.999 & 2.000 \\
8 & 2.000 & 2.000
\end{tabular}
\end{table}

\begin{figure}
\caption{Stability  diagram of the synchronous orbits of low period
         $q=1,2$ in two linearly coupled PFDP's. The stable regions of
         the stationary point, a symmetric 2-periodic orbit, and an
         asymmetric 2-periodic orbit are denoted by SP, SP2, and ASP2,
         respectively. The horizontal (non-horizontal) solid and
         short-dashed boundary lines correspond to synchronous
         (asynchronous) PDB and PFB lines, respectively. For other
         details see the text.
     }
\label{SD1}
\end{figure}

\begin{figure}
\caption{Stability diagram of synchronous asymmetric $2^n$-periodic
         ($n=1,2,3,4,5$) orbits of level $n$ born via synchronous
         supercritical PDB's. ASP2 denotes the stable region of an
         asymmetric orbit of level $1$, and PN also designates the
         stable region of an asymmetric orbit of period N
         (N$=4,8,16,32$). The solid and
         short-dashed boundary lines represent the same as those in
         Fig.~1. The stability diagram starting from the left (right)
         side of the ASP2 is shown in (a) [(b)]. Note its treelike
         structure. See the text for other details.
     }
\label{SD2}
\end{figure}

\begin{figure}
\caption{ Plots of the asynchronous residue $R_{2,n}(A^*_1,c)$ versus
     $c$ near the zero-coupling critical point for $n=4,5,6$.
     }
\label{AR1}
\end{figure}

\begin{figure}
\caption{ Plots of the asynchronous residue $R_{2,n}(A^*_1,c)$ versus
$c$ near the rightmost critical line in Fig.~2(a) for $n=5,6,7$.
     }
\label{AR2}
\end{figure}

\begin{figure}
\caption{Maximum asynchronous Lyapunov exponent $\sigma_{2,1}$ of the
         synchronous quasiperiodic orbit near the rightmost critical
         line in Fig.~2(a). The values of $\sigma_{2,1}$ at both ends
         of the rightmost critical line are zero, which are denoted
         by solid circles.
     }
\label{ALYP1}
\end{figure}

\begin{figure}
\caption{Maximum asynchronous Lyapunov exponents $\sigma_{2,1}$ of the
         synchronous quasiperiodic orbit near both (a) the left end
         and (b) the right end of the rightmost critical line in
         Fig.~2(a).
         Here $\epsilon = c - c^*$ $(c^*=c_l$ or $c_r$). Note that
         $\sigma_{2,1}$ varies linearly with respect to $c$ near both
         ends.
     }
\label{ALYP2}
\end{figure}

\begin{figure}
\caption{ (a) Bifurcation diagram (plot of $x^*$ versus $A$) in the
  vicinity of the first resurrection of the stationary point with
  $x^*=0$; $x_1=x_2 \equiv x^*$ for a synchronous orbit. Here $q=1(2)$
              denotes the period of a synchronous orbit, born via
              supercritical PFB (subcritical PDB). The solid and
              short-dashed lines also designate stable and unstable
              orbits, respectively.
          (b) Second stability diagram of synchronous orbits near the
              $c=0$ line. Here SP, ASP1, and PN denote the stable
              regions of the stationary point, an asymmetric orbit of
              period $1$,
              and an asymmetric N-periodic (N$=2,4,8,16)$ orbit,
              respectively. The solid and short-dashed boundary lines
              also represent the same as those in Fig.~1. For other
              details see the text.
     }
\label{BDSD1}
\end{figure}

\begin{figure}
\caption{ (a) Bifurcation diagram (plot of $x^*$ versus $A$) in the
vicinity of the second resurrection of the stationary point with
$x^*=0$; $x_1=x_2 \equiv x^*$ for a synchronous orbit. Here $q=1(2)$
denotes the period of a synchronous orbit, born via
subcritical PFB (supercritical PDB). As in Fig.~7(a), the solid
and short-dashed lines also designate stable and unstable
orbits, respectively.
(b) Third stability diagram of synchronous orbits near the $c=0$ line.
Here SP, SP2, ASP2, and PN (N$=4,8,16)$ denote the stable
regions of the stationary point, a symmetric orbit of period
$2$, an asymmetric $2$-periodic orbit, and an
asymmetric orbit with period N (N$=4,8,16)$ orbit, respectively.
The solid and short-dashed boundary lines
also represent the same as those in Fig.~1. For other
details see the text.
     }
\label{BDSD2}
\end{figure}

\begin{figure}
\caption{ Stability diagrams of synchronous orbits near the $c=0$ line
for the cases of (a) the quadratic and (b) cubic couplings.
Here SP2, ASP2, and PN (N$=4,8$) denote the stable regions of
a symmetric orbit of period $2$, an asymmetric $2$-periodic
orbit, and an asymmetric orbit with period N, respectively.
         }
\label{NCSD}
\end{figure}

\begin{figure}
\caption{ Stability diagram of synchronous orbits in four
linearly-coupled PFDP's. Each stable region is bounded by its
solid boundary
curves. For a synchronous orbit of period $q$, the PDB (PFB) curve
of the mode with index $j$ is denoted by a symbol $q^{PD(PF)}_j$.
The stability diagram starting from the left (right) side
of a $2$-periodic orbit is shown in (a) [(b)]. For other details
see the text.
     }
\label{MCSD}
\end{figure}


\begin{references}
\bibitem{Kook2} Without giving any explicit numerical values of the
 scaling factors, the authors of the paper \cite{Kook1} insisted
 that the critical behaviors of the coupled oscillators are the
 same as those of the coupled 1D maps, based on the stability
 diagram of orbits with only period $2$ and $4$ (see Figs.~1
 and 2 in \cite{Kook1}).
\bibitem{Rem} Unlike the case of the first period-doubling transition
to chaos with only one ``stability tree'' consisting of connected
stability regions of synchronous orbits of all levels, several separate
stability
trees are found in the $A-c$ plane for the case of the second (third)
period-doubling transition to chaos. Only the stability tree starting
from the stability region of the stationary point including the $c=0$
line is shown in Fig.~\ref{BDSD1}(b) [\ref{BDSD2}(b)], whose structure
is essentially the same as that in Fig.~\ref{SD2}(a). However, other
separate stability trees grow like chimneys without $U$-shape
branchings, as in Fig.~\ref{SD2}(b), where the critical behaviors are
also the same as those in the $C$-route for the first period-doubling
transition case.
\end{references}
\end{document}